# Assessing the influence of one astronomy camp over 50 years

**Hannah S. Dalgleish***

*Former President of IWA*
*Astrophysics Research Institute, Liverpool John Moores University, Liverpool, UK*
hannah@iayc.org

**Joshua L. Veitch-Michaelis**
*Vice President of IWA*
*Natural Sciences and Psychology, Liverpool John Moores University, Liverpool, UK*

*The International Astronomical Youth Camp has benefited thousands of lives during its 50-year history. We explore the pedagogy behind this success, review a survey taken by more than 300 previous participants, and discuss some of the challenges the camp faces in the future.*

Summer camps have been enriching the lives of young people for over 100 years. There are now numerous camps dedicated to space and astronomy, such as Space Camp (1982), AstroCamp (USA, 1988), Space School (UK, 1989) and the European Space Camp (1996). Established in 1969, the International Astronomical Youth Camp (IAYC) predates these camps by more than a decade and, as far as we are aware, is both the oldest and longest-running event of its kind.

The first IAYC was held in Schmallenberg, West Germany in 1969, and was organised primarily by amateur astronomer Werner Liesmann[1]. The camp hosted 40 participants who were lectured in astronomy and took part in project work; reportedly, it was a great success and 90 people attended the following year[2]. Ten years later, in 1979, the International Workshop for Astronomy (IWA) was founded as the formal organisation behind the camp. The current statutes of IWA list two aims:

- *Promotion of international collaboration and agreement, particularly for astronomical youth work*
- *To spread astronomical knowledge and to teach young people to work scientifically on their own*

This year (2019) marks the 50th anniversary of the IAYC, and the 55th camp since 1969 (some years had additional camps). Although it now lasts three weeks instead of one, the format of the IAYC is largely unchanged; for a detailed description of the camp see Dalgleish[3]. So far, IAYCs have been held in 15 countries (Figure 1) and have hosted 3,468 participants (of whom we estimate at least 1,700 to be unique), aged 16-24, from 81 nationalities worldwide (Figure 2).

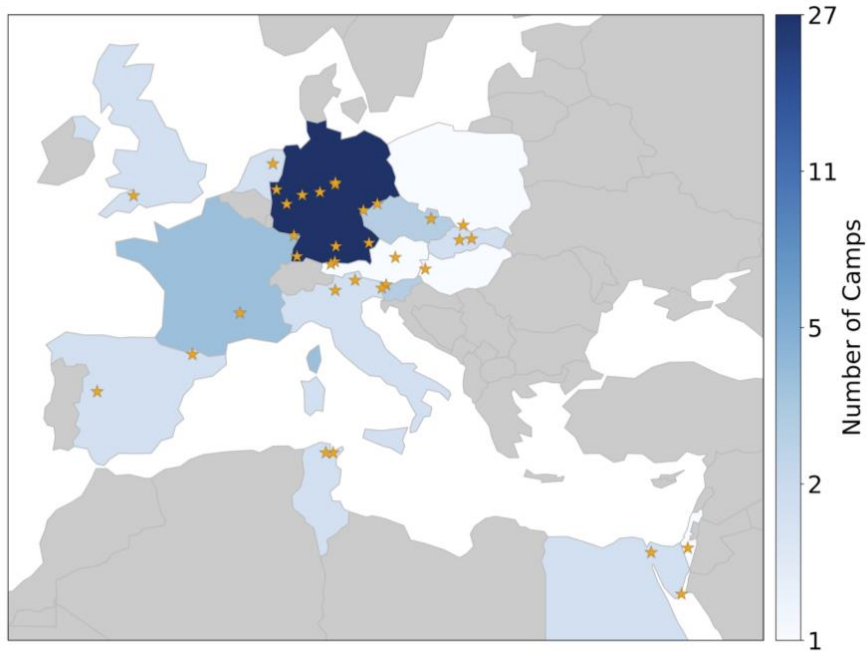

**Figure 1 | Locations of the 55 IAYCs since 1969.** Orange stars represent camphouse locations, many have been used more than once. The IAYC originated in Germany and has been held there 27 times.

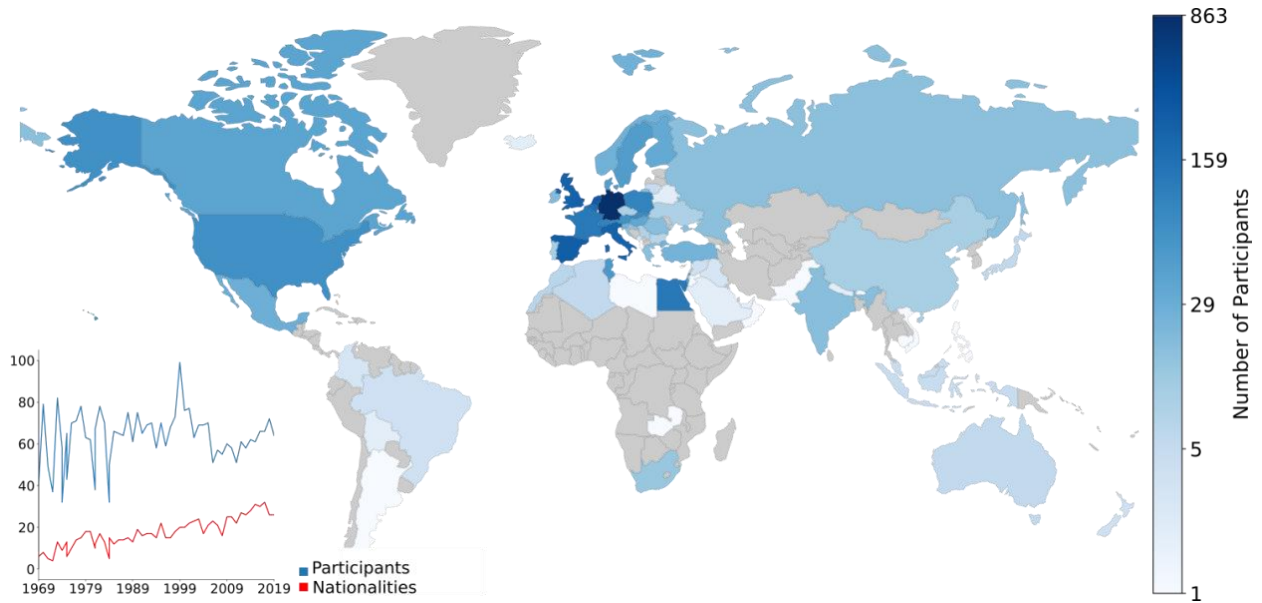

**Figure 2 | Distribution of participant nationalities across all IAYCs.** Colours are on a log scale, grey indicates no participants. **Inset:** Number and nationalities of participants for each camp between 1969 and 2019. The camp with peak participation coincided with the 1999 [solar eclipse](#) over Europe.

## Pedagoagoch approach

The IAYC continues to attract young people from all over the world, and up to 50% of each camp is filled by returning participants. This success can be attributed to the IAYC's pedagogical approach; instead of offering lectures or timetabled classes like other science camps, the IAYC encourages informal learning through guided, astronomy-related, project work (similar to project-based or cooperative learning e.g. ref. 4).

Participants work on a research project over the course of the three weeks, culminating in a final report (e.g. ref. 5). The goal of the camp is not to produce novel scientific results, but to bridge cultural barriers while fostering curiosity and intuition for how research is carried out. The IAYC often hosts participants from non-STEM backgrounds who are not interested in becoming astrophysicists, but who simply enjoy stargazing.

Working groups are central to the IAYC. Each group, of around eight people, has a leader — a former participant who is often an undergraduate or postgraduate student with experience in astronomy or another relevant subject. For many participants, the IAYC is their first experience working independently or even away from home. We note that this structure reflects the idea of 'affinity spaces'[6]: the participants come together due to a shared interest in astronomy, advancing knowledge through engaging and working with others from different backgrounds.

Leaders provide a range of project ideas that can be adapted to the participants' previous experience, and are both engaging and relevant for modern-day astrophysics. Observational and theoretical projects are most common, with many group projects involving programming — machine learning projects have also been recently offered. Instrumentation-based work is often possible, and requires more careful planning. Participants are also encouraged to suggest their own ideas, which may draw upon their own research. By the end of the camp, final reports are written in LaTeX, a skill useful to real-world publication.

Determining the success of a project is subjective. Sometimes objective goals can be defined, such as successfully imaging a certain Messier object. For other more exploratory projects, such as investigating the influence of astronomy on science fiction, it is less straightforward. Arguably, success is best measured by participant satisfaction; while a project may be considered to be a "failure" (e.g. building a radio telescope but unsuccessfully making observations), the participants will have likely learned a great deal in the process. This allows participants to experience failure as a necessary part of the scientific method.

> *"The camp prepared me well for college, taught and motivated me to gain news skills, gave me confidence in science, as well as a strong belief that I can be a successful female physicist."*

> *"[The camp] helped me think critically and independently. It helped me understand science better. I learned how to start a project, follow through and write up a presentation on it."*

## The legacy of the IAYC

Despite not being the main focus of the camp, there are cases where project work has influenced participants' future research (for example, ref. 7 cites an IAYC project) or where friendships have led to fruitful collaborations — at least two cases resulted in a *Nature* paper[8,9]. Searching the *NASA Astrophysics Data System* (ADS), we found more than 25 other *Nature* papers authored or co-authored by past participants of the IAYC, in addition to ~2750 refereed papers in other astronomical journals.

Using data from Kamphuis & van der Kruit[10], we can make a rough comparison between the impact of IAYC astronomers (that are known to us) and a subset of members of other astronomical societies (Table 1). Our sample is made up of previous participants whose 1st peer-reviewed, first-author paper was published in 2015 and before; interestingly, 16 out of 26 astronomers were previous leaders at the camp. For the citation count, we use the number of references to that article in all bibliographic publications, both refereed and non-refereed, as in ref. 10. Although our sample size is an order of magnitude smaller, the data suggest that IAYCers who continue in full-time astronomy careers are at least as successful as astronomers who did not attend the camp.

|  | IAYC (postdocs and above) | American Astronomical Society (full members) | International Astronomical Union (active members) |
| --- | --- | --- | --- |
| Sample size | 38 | 172 | 193 |
| First-author papers per year | 0.8 | 0.5 | 0.6 |
| First-author citations per year | 39 | 8 | 9 |

**Table 1 | Comparison of first-author papers and citations per year between different groups of astrophysicists.** Data obtained via ADS and Kamphuis & van der Kruit[10].

Additionally, the camp has seen some prolific astronomers. For example, Robert McNaught discovered 82 comets and 483 minor planets, and is quoted as being "the world's greatest comet discoverer"[11]. Erich Karkoschka also made a significant contribution to astronomy; he discovered Perdita, one of Uranus' moons[12] and has an asteroid named in his honour (30786 Karkoschka). The famous Dutch popular-science journalist, Govert Schilling, equally has an asteroid named after him (10986 Govert).

Meanwhile, other participants have been inspired to establish new events such as the Meteor Seminars[13] (now the International Meteor Conferences), or the International Science Engagement Camp. While it is encouraging to see this evidence, there is still much to learn about the camp's influence on the participants as a whole. For this reason, we undertook a survey of those who had previously attended the IAYC.

## Participants survey

The IAYC survey was launched on 16th August, 2017, with preliminary results presented at the *Communicating Astronomy with the Public* conference, 2018[14]. The following results represent an expanded follow-up survey totalling 307 respondents of 58 different nationalities; ages range from 16 to 79, most of whom would now be ineligible to return to the camp. 119 respondents identified as female, 181 as male and the remainder as "other" or declined to answer. The vast majority of respondents (260) identified as white. 215 people attended multiple times and responses cover every year of the camp's existence (Figure 3).

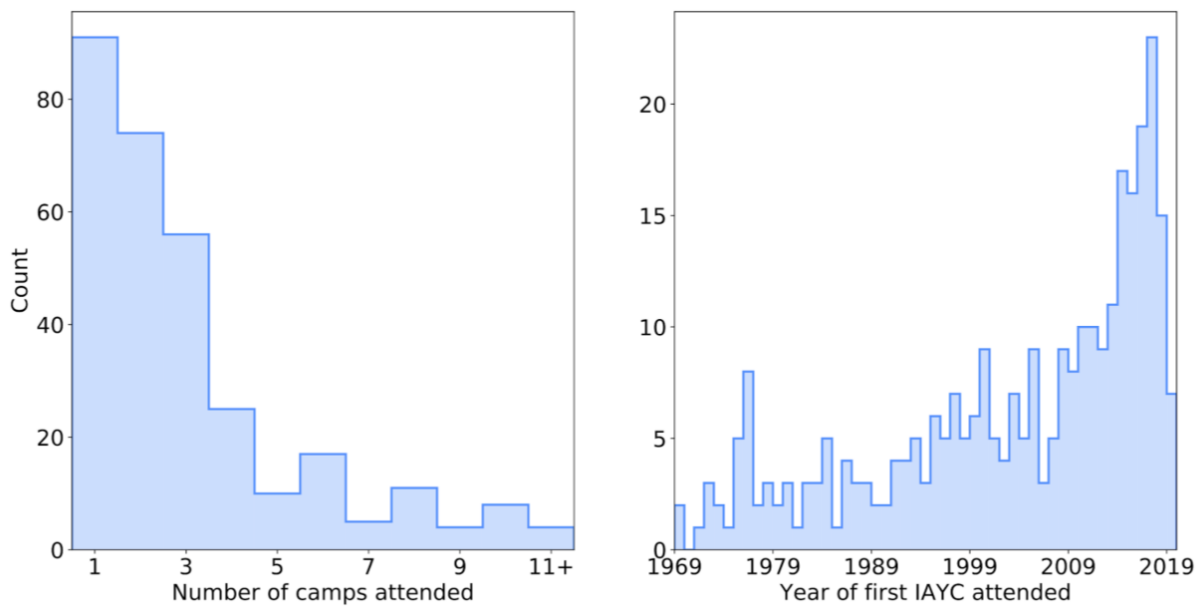

**Figure 3 | Number of IAYCs attended by survey respondents (left) and when they first attended (right).** Those who participated nine times or more were all leaders. One of the respondents did attend the IAYC in 1970, but it was not their first camp.

We asked a series of 54 questions relating to the respondents' backgrounds, and how the IAYC affected them academically and personally. Here we focus on a few key educational and personal aspects of the data that has been included with the respondents' consent.

## Educational impact

To explore the camp's influence on career choice, we asked respondents to provide their current occupation as well as information on their astronomical background prior to attending the camp (Table 2). While this data does not indicate a causal relationship, previous studies have found that young people were more likely to report a career interest in science and technology as a result of participating in a science camp[15].

|  | Yes | No | Maybe |
|---|---|---|---|
| Studied astrophysics prior to IAYC | 62 | 240 | - |
| Astronomy as a hobby prior to IAYC | 238 | 69 | - |
| IAYC influenced career choice in some way | 145 | 81 | 81 |
| Chose to study or continue studying astrophysics due to IAYC | 29 | - | - |

**Table 2 | Astronomical background prior to IAYC and influence on education/career choice.**

For those who aged 25-65 (195 respondents), we found that 14.4%, 37.4% and 46.7% held Bachelor, Master and PhD degrees as their highest degree, respectively. For comparison, in 2018, 32.3% of adults aged 25-64 in the 28 EU member states had attained some form of tertiary education ([Eurostat](#) database, 2019). Additional results from the survey (with a smaller sample size of 100) reveal that 77% of participants have parents educated to at least Bachelor level, implying that the majority of participants are from higher socio-economic backgrounds.

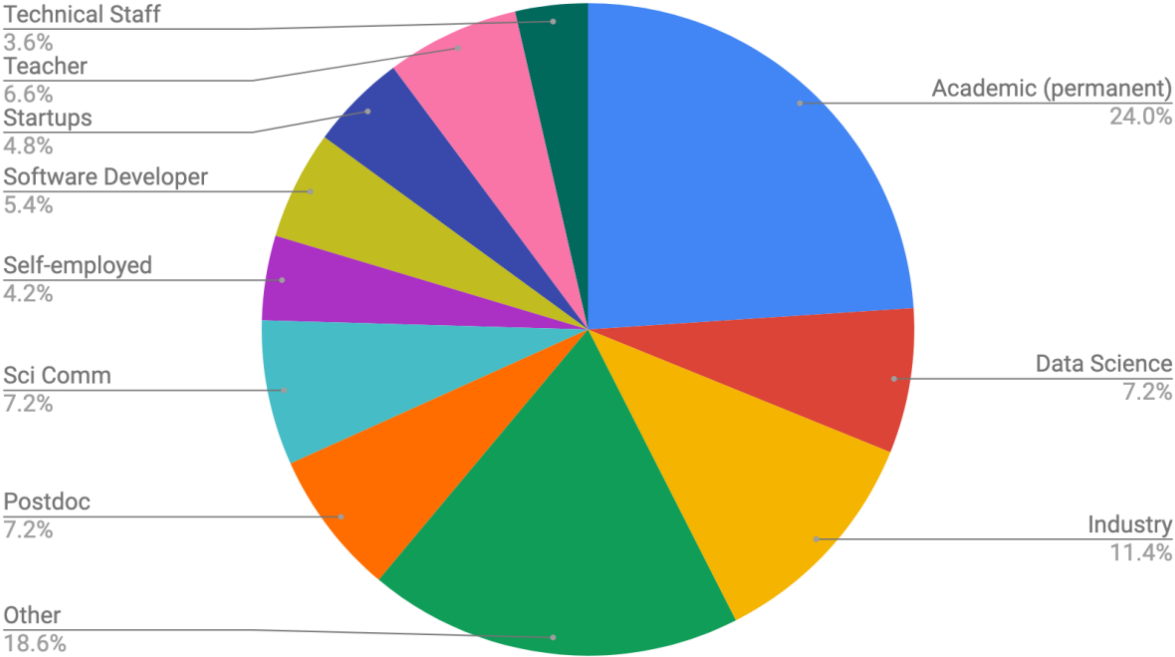

**Figure 4 | Careers of non-students aged 25-65.** 'Other' includes those who are retired or unemployed and careers such as medicine or politics.

Of the non-students in the 25-65 age range (167 respondents), 31.2% reported working in academia in some form (Figure 4) which includes 50.0% within astronomy and 30.8% in other STEM-related fields. Of the students (129 respondents; across all age ranges), 27.1% are studying astronomy and 49.6% are studying a STEM-related subject. Some noted in further detail:

> *"The IAYC acted as an incredible motivator, a catalyst to pursue a career in science. It made me feel that this is what I wanted to do. Every year, coming back from the camp, I was thrilled to learn and dive deeper into my subject."*

> *"The IAYC showed me it was possible to pursue astronomy and astrophysics and make a career out of it. I didn't really realise I could just do this very specific area of physics with my life. A career in astronomy/astrophysics seemed possible after my first camp."*

For some who are first generation academics, the camp appears to have had a profound effect: the camp motivated them to diverge away from socio-cultural norms and pursue careers in science.

> *"I am one of eight children... None of my siblings went to college from school. My trip to the IAYC, sponsored by a national scientific organisation, confirmed and inspired my love of astrophysics, making me determined to pursue it as a career. I subsequently completed an undergraduate degree, Masters and a PhD in mathematics, with a specialty in general relativity."*

> *"For [me] the international, gender-balanced character of the camp was a real eye opener. It made me look beyond my narrow 18 year old horizon. The world opened for me. This feeling of being a global citizen capable of becoming a scientist came straight from the IAYC. The camp gave me the confidence I needed to pursue an international career in science."*

Several other respondents mentioned that they were already planning to pursue a career in science, while others shared that the camp helped them to realise that astronomy wasn't for them. There were also comments on how the IAYC broadened horizons and increased confidence when applying for future roles, especially due to an increased proficiency in English. Many were inspired to travel and pursue opportunities abroad (84.4%), to collaborate internationally (77.2%), and we found that 35.5% now live in a different country compared to where they are from. Participants also noted that engaging with such a diverse group of people was useful for networking. Of the more recent participants, most reported the lack of internet access at the camp as a positive influence, since it encouraged social interactions as well as independent problem solving.

## Personal impact

To further explore the impact on the participants' personal lives, we asked how many of the respondents have remained in contact. The prolonged friendships established during the camp and beyond are significant — for several people, these friendships have lasted a lifetime. We found that 54.4% have taken part in reunions, and 74.9% are in contact on a monthly basis or more, demonstrating the long-term friendships that emerge. This is further shown by the number of participants who are now married or in partnership (7.8%) and have children together (3.3%). Peer relationships were also prominent in the study by Fields[16], based on an astronomy camp in the USA.

We also asked whether the camp had been helpful for those who identified as a member of a minority group (33.9%) or had experienced impostor syndrome (44.0%) — of whom 53.8% (35.6% maybe) and 33.3% (41.5% maybe) felt that the camp had helped them to reach their goals, respectively.

> *"[The IAYC] opened a whole new world of friends, travels, experiences. It made me a much more international person and in some way allowed me to be more confident."*

> *"[The IAYC] changed in every way my social life, I went from being a very shy person to an outgoing, open and kind one. It also taught me the humility and the perspective of life that comes with learning astronomy."*

Respondents reported a positive effect having attended the IAYC, from perceived self-awareness and confidence to increased social inclusion. As our sample is not representative of all previous participants, we are cautious not to extrapolate; responses from non-STEM/astronomers or for whom the camp had a negative influence are few. Analysis of the data is ongoing; we plan to publish the results in full on [the IAYC website](the IAYC website).

## IAYC in the future

The survey has shown that the format of the camp is highly effective for astronomy and STEM outreach, yet work is ongoing in order to remain relevant academically and culturally in the future. In the following, we describe some of the challenges IWA currently faces.

### Camp accessibility and diversity

The overall gender and ethnic diversity of the camp is promising. Recent camps have hosted roughly equal numbers of female and male participants[14], and we have observed an increase in those identifying their gender as neither female nor male. The number of nationalities represented has been increasing almost continually since the camp started (Figure 2, inset), however, a major issue is a lack of representation from developing countries, particularly in Africa and South America. This is observed in camp applications as well as participation demographics. We attribute this to several factors including location, cost and discoverability.

Travel and visa costs are significant for applicants outside Europe, thus it is unsurprising that the majority of applicants are European; while the camp fee is set to be approximately at-cost, it is unaffordable for young people in developing nations. We raise funding for a limited grant program towards fees, but receive more requests for funding than we can satisfy (by a factor of ~7). Other organisations have possibly mitigated this problem by offering variable fees determined by participant nationality and subsidising costs for disadvantaged applicants (e.g. the International Association for Physics Students).

Advertising poses another limitation which is predominantly carried out via word-of-mouth, contacts in astronomical societies, and high schools/university departments (often restricted to Europe). We increasingly rely on "organic" discovery via web searches, and the camp currently ranks highly in common search engines (e.g. Google) for English phrases like "astronomy camp". Therefore, applications are likely biased towards those already interested in astronomy or who know previous participants.

A straightforward solution is that more, local, events are required to support increasing numbers of young people interested in astronomy around the world. This issue of under-representation in astronomy is well known in the wider community, for example, the International Astronomical Union (IAU) designates a list of countries that should be targeted for outreach and development (e.g. ref. 17).

### Dealing with oversubscription

Applicants are evaluated based on a short motivation statement, rather than academic performance. Historically, places were offered on a first-come, first-served basis until the camp was full. This is no longer possible and IWA is now evaluating strategies to cope with the camp being increasingly oversubscribed (i.e. 179 applicants for 64 places in 2019) whilst maintaining diversity within the cohort. This process is the result of several years of iterative improvement.

To limit the effects of unconscious bias, multiple reviewers make a blind, subjective, evaluation of participants based on a motivational statement and limited supporting information. However, blinding may lead to other selection biases, for example how fluent the applicant is in written English. Recently, we tested an algorithm similar to Entrofy[18] which frames applicant selection as an optimisation problem; event organisers can identify diversity criteria among applicants such as gender or nationality. We are still experimenting with this approach, combined with our current selection process, but have found that this still requires significant human oversight and needs further iteration.

## Acknowledgements

Organising the IAYC is a team endeavour. The camp would not exist were it not for the considerable efforts of more than 120 volunteers, and thus we would like to acknowledge all current and past leaders who have made the camp possible since 1969, as well as all the participants who agreed to fill out the survey. We would also like to thank the current members of IWA who provided comments on the final draft as well as N. Mórocza for her comments and advice on survey interpretation.